\documentclass[aps,showpacs,showkeys,preprintnumbers,notitlepage]{revtex4-1}

\usepackage{epsfig}
\usepackage{multirow}
\usepackage{amsmath}
\usepackage{amsfonts}
\usepackage{amssymb}
\usepackage{graphicx}
\usepackage{color}
\usepackage{ulem}

\usepackage[colorlinks=true,citecolor=blue,linkcolor=blue]{hyperref}

\begin{document}

\title{Searching for cavities of various densities in the Earth's crust\\with a low-energy $\bar{\nu}_e$ $\beta$-beam}

\author{C.A.~Arg\"{u}elles}
\email{arguelles@wisc.edu}
\altaffiliation{{\bf Present institution:} Dept.~of Physics and Wisconsin IceCube Particle Astrophysics Center, University of Wisconsin, Madison, WI 53706, USA}
\author{M.~Bustamante}
\email{bustamanteramirez.1@osu.edu}
\altaffiliation{{\bf Present institution:} Center for Cosmology and AstroParticle Physics (CCAPP), The Ohio State University, 191 W.~Woodruff Ave., Columbus, OH 43210, USA}
\author{A.M.~Gago}
\email{agago@pucp.edu.pe}
\affiliation{Secci\'on F\'isica, Departamento de Ciencias, Pontificia Universidad Cat\'olica del Per\'u, Apartado 1761, Lima, Peru}

\date{August 17, 2015}


\begin{abstract}
We propose searching for deep underground cavities of different densities in the Earth's crust using a long-baseline $\bar{\nu}_e$ disappearance experiment, realized through a low-energy $\beta$-beam with highly-enhanced luminosity. We focus on four cases: cavities with densities close to that of water, iron-banded formations, heavier mineral deposits, and regions of abnormal charge accumulation that have been posited to appear prior to the occurrence of an intense earthquake. The sensitivity to identify cavities attains confidence levels higher than $3\sigma$ and $5\sigma$ for exposures times of $3$ months and $1.5$ years, respectively, and cavity densities below $1$ g cm$^{-3}$ or above $5$ g cm$^{-3}$, with widths greater than $200$ km. We reconstruct the cavity density, width, and position, assuming one of them known while keeping the other two free. We obtain large allowed regions that improve as the cavity density differs more from the Earth's mean density. Furthermore, we demonstrate that knowledge of the cavity density is important to obtain O(10\%) error on the width. Finally, we introduce an observable to quantify the presence of a cavity by changing the orientation of the $\bar{\nu}_e$ beam, with which we are able to identify the presence of a cavity at the $2\sigma$ to $5\sigma$ C.L.
\end{abstract}


\maketitle

\section{Introduction}

One of the most interesting findings in particle physics in the last two decades is the fact that neutrinos have non-zero masses. As a consequence, they can transform periodically between different flavors as they propagate. This quantum-mechanical phenomenon is known as neutrino oscillations \cite{Fukugita:2003en,GonzalezGarcia:2002dz}, and has been confirmed by overwhelming experimental evidence (see, {\it e.g.}, Refs.~\cite{Balantekin:2013tqa,Bellini:2013wra} and references therein). The present and future experimental efforts are focused on the precision measurement of the neutrino oscillation parameters and also on searches for hints of physics beyond the Standard Model in the neutrino sector. The possibility of having detailed knowledge of these parameters, together with advances in the experimental techniques in neutrino production and detection, has created an appropriate scenario for proposing neutrino technological applications such as neutrino communication
\cite{MINERvA:2012en,Huber:2009kx}, neutrino tomography of the Earth \cite{DeRujula:1983ya,Winter:2006vg}, and others \cite{Christensen:2013eza}.

These technological applications profit from different neutrino properties. For instance, neutrino communication is interesting due to the fact that the neutrino interacts weakly with matter, making it possible to establish a link with a receiver that is inaccessible by conventional means, {\it i.e.}, electromagnetic waves, which are either damped or absorbed by the intervening medium. On the other hand, the proposed neutrino tomography of the Earth's interior relies on the sensitivity of the oscillation probability to the matter density along the neutrino path. Inspired by this idea, some studies have been made on the possibility of using neutrinos to search for regions of under- and over-density compared to the average density of the Earth's crust; notably in the context of petroleum-filled cavities, employing either a superbeam \cite{Ohlsson:2001fy} or the flux of $^7$Be solar neutrinos \cite{Ioannisian:2002yj}, and of electric charge accumulation in seismic faults prior to earthquakes \cite{Wang:2010cb}, employing reactor neutrinos.

In this letter, in comparison with previous works, we have made a more detailed analysis by including a better description of the experimental setup, such as the neutrino flux description and a likelihood analysis. In this more realistic framework, we have calculated, for the first time, the sensitivity to the detection of an underground cavity as a function of its parameters (position, width, and density). Our experimental arrangement considers a long-baseline ($1500$ km) neutrino disappearance experiment using a low-energy ($5$\---$150$ MeV) $\beta$-beam \cite{Volpe:2006in} source, for which we have studied four different cavity scenarios: water-like density \cite{Pearson2014}, an iron-banded formation \cite{Morey:1983}, a heavier mineral deposit, and a zone of seismic faults \cite{Pulinets:2004}. Finally, another additional innovative idea is the introduction of a movable configuration of beams and detectors.

\begin{figure}[t!]
 \begin{center}
  \scalebox{0.35}{\includegraphics{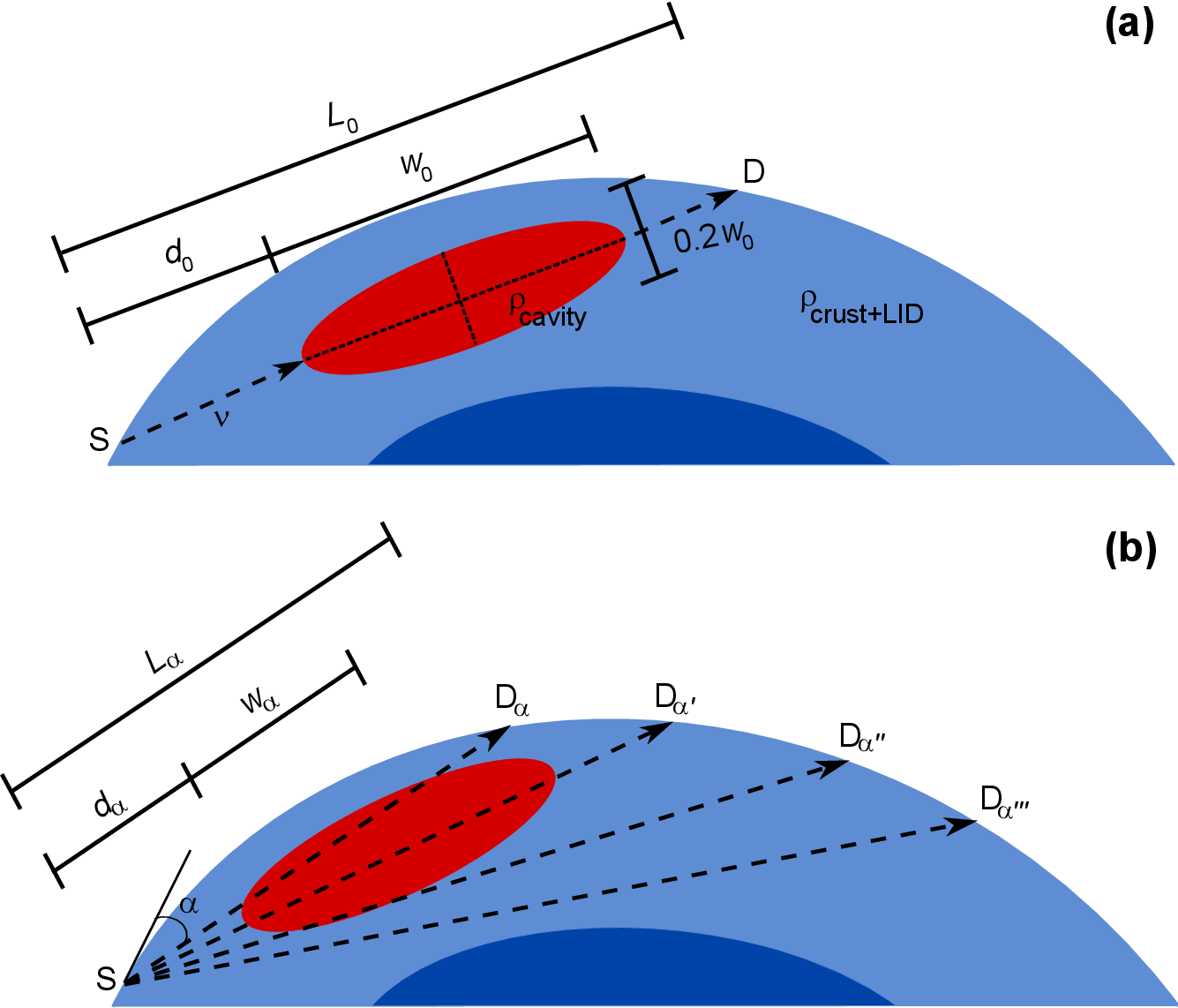}} 
  \caption{\label{FigCavityConfiguration}(Color online) Cavity and neutrino beam with a fixed (a) and different orientations (b).}  
 \end{center}
\end{figure}

\section{Neutrino propagation in matter}

The probability amplitudes for the neutrino flavor transitions $\bar{\nu}_e \rightarrow \bar{\nu}_\beta$ can be arranged in a column vector $\Psi_e = \left( \psi_{ee} ~~ \psi_{e\mu} ~~\psi_{e\tau} \right)^T$ which evolves according to $i d\Psi_e/dx  = H \Psi_e$, where $x$ is the distance traveled since creation and the effective Hamiltonian in the flavor basis is given by $H\left(x\right) = 1/\left(2E_\nu\right) U^\dagger \text{diag}\left(0,\Delta m_{21}^2,\Delta m_{31}^2\right) U + A\left(x\right)$, with $E_\nu$ the neutrino energy, $\Delta m_{21}^2$ and $\Delta m_{31}^2$ squared-neutrino mass differences, and $U$ the lepton mixing matrix \cite{Nakamura:2010zzi}. The matter effects are encoded in the matrix $A\left(x\right) = \text{diag}\left(-\sqrt{2} G_F N_e\left(x\right),0,0\right)$, where $N_e \left(x\right)= y_e \rho\left(x\right) N_\text{Av}$ is the electron number density, with $\rho$ the matter density, $N_\text{Av}$ Avogadro's number, and $y_e = 0.494$ the average 
electron fraction in the Earth's crust \cite{Wang:2010cb}. Our results have been obtained by numerically solving the evolution equation described above, where we have fixed the values of the squared-mass differences and angles to the best-fit values of  Ref.~\cite{Schwetz:2014qt} and set the CP phase to zero. It is important to point out that given that we will study only the $\bar\nu_e$ survival probability the value of the CP phase is not important; furthermore, since this probability is driven by the solar scale $\Delta m^2_{21}$, the sign of $\Delta m^2_{31}$ is also not relevant. 

While our results have been obtained by numerically solving the evolution equation, in order to gain a qualitative understanding of its behavior, we will refer in our discussion to the following well-known approximation of the oscillation probability \cite{Peres:2009xe}: $P_{\bar{\nu}_e\to\bar{\nu}_e}^{3\nu} = P_{\bar{\nu}_e\to\bar{\nu}_e}^{2\nu} \cos^4 \theta_{13} + \sin^4 \theta_{13}$ (with $A \rightarrow A \cos^2 \theta_{13}$), where $P_{\bar{\nu}_e\to\bar{\nu}_e}^{2\nu}$ is the two-flavor slab approximation calculated for a piecewise constant density profile \cite{Akhmedov:1998ui,Ohlsson:1999um} made up of three matter layers, corresponding to the sections of the Earth's interior that are traversed by the neutrino before entering the cavity, inside of it, and after exiting it. It is described by $P_{\bar{\nu}_e\to\bar{\nu}_e}^{2\nu} = |\left[\mathcal{U}\right]_{11}|^2$, where $\mathcal{U} = \mathcal{U}_3 \times  \mathcal{U}_2 \times \mathcal{U}_1$ is the probability amplitude of the $\bar{\nu}_e$ surviving the traversal of the three layers, $\mathcal{U}_k = \cos \phi_k - i (\vec{n}_k \cdot \
\vec{\sigma}) \sin \phi_k$, $\vec{\sigma}$ is the vector of Pauli matrices, and $\vec{n}_k = (\sin 2 \theta_{M_k}, 0, -\cos 2 \theta_{M_k})$, with $\theta_{M_k}$ the value of the $\theta_{12}$ mixing angle modified by matter effects in the $k$-th slab. The frequencies $\phi_k$ are given by $\Delta m^2_{M_k} x_k/(4E_\nu)$, where $x_k$ is the width of the $k$-th slab, and $\Delta m^2_{M_k}$ is the matter-modified value of $\Delta m_{21}^2$. Since the first and third slabs correspond to the crust, while the second one corresponds to the cavity, we set $x_1 = d$ (distance from the surface to the first point of contact of the beam with the cavity), $x_2 = w$ (width of the cavity traversed by the beam), and $x_3 = L_0-d-w$ (with $L_0$ the total baseline of the beam). Our numerical computation of the three-flavor oscillation probability and this approximation are in reasonable agreement, to within $\sim 1\%$.

\textit{Location and shape of the cavity inside the Earth}.\--- We have assumed the existence of a cavity of uniform density $\rho_\text{cavity}$ located within the Earth's crust, itself of density given by the Preliminary Reference Earth Model (PREM) \cite{Dziewonski:1981xy}. Together, the ocean, crust, and LID (the low velocity zone, which is the main part of the seismic lithosphere) layers of the PREM have a depth of up to $80$ km and an average density of $\langle\rho_\oplus\rangle =  3.3$ g cm$^{-3}$. Interesting geological features such as porous rock cavities, mineral deposits, and seismic faults lie in the crust and LID layers. Therefore, we have supposed that no part of the cavity is below $80$ km. The cavity itself has been modeled as an ellipsoid, and we have studied an elliptic cross section of it, with major axis length $w_0$ and minor axis length $0.2 w_0$, as shown in Fig.~\ref{FigCavityConfiguration}a. This choice of shape constitutes a toy model of a real cavity, which might, of course, have a more complicated shape. Underground oil reservoirs and aquifers, in particular, have a flatter shape. However, our choice of an ellipsoid serves to more easily test the capability of our proposed method. After fixing the value of the source-detector baseline, $L_0$, we position the neutrino source (S) on the Earth's surface. Since the adopted density profile of the Earth is radially symmetric, we can place the source at any position on the surface. In order to specify the location of the cavity, we set the distance $d_0$ measured along the baseline from the source to the cavity's surface.

\section{Low-energy $\beta$-beams}

There is currently a proposal to use a pure, collimated beam of low-energy $\bar{\nu}_e$ generated by means of the well-understood $\beta$ decay of boosted exotic ions \cite{Volpe:2006in} and detected through $\bar{\nu}_e + ^{12}$C $\rightarrow e^+ + ^{12}$B \cite{Samana:2010up}. Our $\beta$-beam setup contemplates an ion storage ring of total length $l_\text{tot} = 1885$ m, with two straight sections of length $l_\text{straight} = 678$ m each \cite{Balantekin:2005md}. Inside the ring, $^6$He ions boosted up to a Lorentz factor $\gamma = 25$ decay through  $^6_2$He$^{++} \rightarrow ~ ^6_3$Li$^{+++} + e^- + \bar{\nu}_e$ with a half-life $t_{1/2} = 0.8067$ s. Ion production with an ISOLDE technique \cite{Autin:2002ms} is expected to provide a rate of ion injection of $g = 2 \times 10^{13}$ s$^{-1}$ for $^6$He; we have introduced a highly optimistic $5000$-fold enhancement of this value, which has not priorly been considered in the literature.

The neutrino flux from the $\beta$ decay of a nucleus in its rest frame is given by the formula \cite{Krane:1998} $\Phi_\text{c.m.}\left(E_\nu\right) = b E_\nu^2 E_e \sqrt{E_e^2-m_e^2} F\left(\pm Z,E_e\right) \Theta\left(E_e-m_e\right)$, \;\; where \;$b = \ln{2} / \left(m_e^5 ft_{1/2}\right)$, with $m_e$ the electron mass and $ft_{1/2}=806.7$ the comparative half-life. The energy of the emitted electron is given by $E_e = Q-E_\nu$, with $Q=3.5078$ MeV the $Q$-value of the reaction, and $F\left(\pm Z, E_e\right)$ the Fermi function.

Based on the formalism of Ref.~\cite{Serreau:2004kx}, we have considered a cylindrical detector made of carbon, of radius $R = 4\sqrt{5}~\text{m} \approx 8.94~\text{m}$ and length $h = 100$ m, co-axial to the straight sections of the storage ring, and located at a distance of $L_0 = 1500$ km from it. The integrated number of $e^+$ at the detector, after an exposure time $t$, is calculated as
\begin{equation}\label{equ:DefNumberPositrons}
 N = t g \frac{t_{1/2}}{\ln 2} n h \int d E_\nu \Phi_\text{tot}\left(E_\nu\right) P_{\bar{\nu}_e\to\bar{\nu}_e}\left(L_0,E_\nu\right) \sigma\left(E_\nu\right) ~,
\end{equation}
with $n \approx 6.03 \times 10^{23}$ cm$^{-3}$ the density of carbon nuclei in the detector and $\sigma$ the detection cross section \cite{Samana:2010up}. Since $L_0 \gg l_\text{tot},l_\text{straight},h$, \;we \;can \;write $\Phi_\text{tot}\left(E_\nu\right) \simeq \Phi_\text{lab}\left(E_\nu,\theta=0\right)\left(l_\text{straight}/l_\text{tot}\right)S / \left(4\pi L_0^2\right)$, where $\Phi_\text{lab}$ is the flux in the laboratory frame \cite{Krane:1998}, $\theta$ is the angle of emission of the neutrino with respect to the beam axis, and $S = \pi R^2$ is the detector's transverse area.

\begin{figure}[t!]
 \begin{center}
  \scalebox{0.35}{\includegraphics{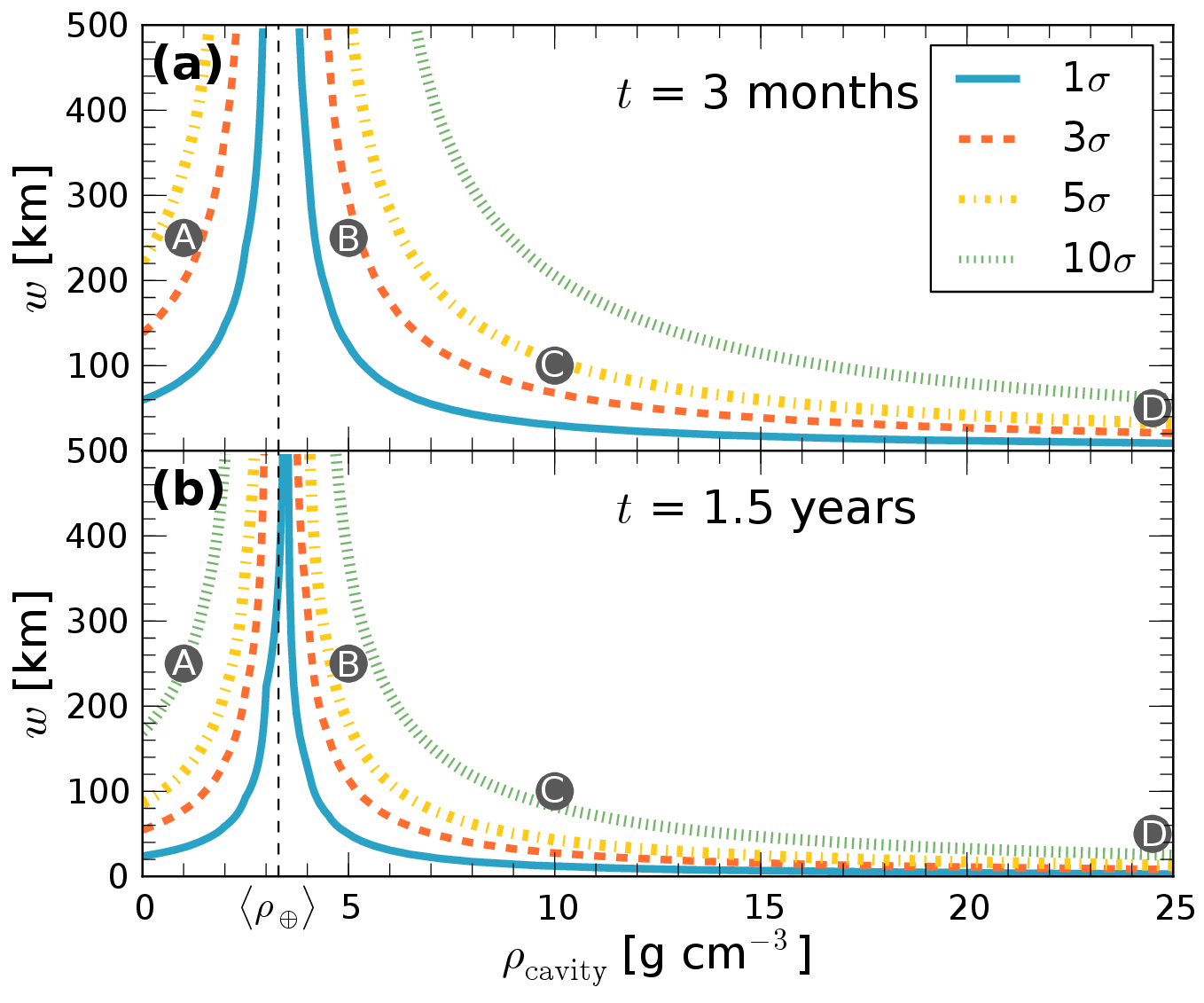}}
  \caption{(Color online) Confidence levels (C.L.s) in the $w$ vs.~$\rho_\text{cavity}$ plane assuming there is no cavity. The four points A\---D are special cases described in the text.}
  \label{Fig-w.vs.rho}
 \end{center}
\end{figure}

\section{Sensitivity to cavities}

We start by assuming that there is no cavity along the baseline $L_0$ and we evaluate the sensitivity to differentiate this situation from the hypothesis that the neutrino beam does traverse a cavity of width $w$, position $d$, and density $\rho$. To do this, we define
\begin{equation}
 \chi^2\left(w,d,\rho \right) = \sum_i \frac{\left[N_i^\text{cav}\left(w,d,\rho\right)-N_i^\text{no-cav}\right]^2}{N_i^\text{no-cav}} ~,
 \label{chi2.nocavity}
\end{equation}
with $N_i^\text{cav}\left(w,d,\rho\right)$ the number of $e^+$, in the $i$-th energy bin, that reach the detector in the case where the beam traverses the cavity, and $N_i^\text{no-cav}$ the corresponding number in the no-cavity case. Given that $\gamma = 25$, the neutrino spectrum extends from 5 to 150 MeV, and we consider bins of 5 MeV. On account of the maximum energy considered, the production of muons, via $\bar{\nu}_\mu + ^{12}$C $\rightarrow \mu^+ + X$, is inhibited, and thus the $\bar{\nu}_e \rightarrow \bar{\nu}_\mu$ channel is not included in this work.  

\begin{table}
 \caption{\label{TblChi0Conts} Minimum value of density $\rho_\text{cavity}$ required to achieve discovery of a cavity of width $w$ centered on a baseline of length $L_0 = 1500$ km, with a statistical significance of $5\sigma$ or $10\sigma$ C.L. (with respect to the no-cavity case), for exposures of $3$ months or $1.5$ years.}
 \begin{center}
   \begin{tabular}{lcccc}
    \hline
    \multirow{2}{*}{Cavity width ($w$)} & \multicolumn{4}{c}{$\rho_{\text{cavity}}\mathrm{[g ~ cm^{-3}]}$} \\
    & \multicolumn{2}{c}{$3$ months} & \multicolumn{2}{c}{$1.5$ years} \\
    & $5\sigma$ & $10\sigma$ & $5\sigma$ & $10\sigma$ \\
    \hline
    50 km  & 18  & $>25$ & 9   & 15 \\
    100 km & 10  & 18    & 6   & 9  \\
    250 km & 6.5 & 9     & 4.5 & 6  \\
    \hline
   \end{tabular}
 \end{center}
\end{table}

Fig.~\ref{Fig-w.vs.rho} shows the sensitivity in the form of isocontours of $\chi^2 = 1\sigma$, $3\sigma$, $5\sigma$, and $10\sigma$, with the hypothetical cavity located at the center of the baseline, so that $d = (L_0-w)/2$ (which is tantamount to $\phi_1 = \phi_3$), and detector exposure times of (a) $3$ months and (b) $1.5$ years. Because of this relation between $d$ and $w$, the analysis can be reduced to two parameters: $\rho$ and $w$ (or $d$). It is observed that the sensitivity improves as $|\rho_\text{cavity}-\langle\rho_\oplus\rangle|$ increases. Also, the statistical significance, for the cavity hypothesis, grows with $w$, given that the cumulative matter density differences are greater. 

In the slab approximation, the behavior of the oscillation probability, which dictates the shape of the isocontours, is dominated by the $\phi_2$ frequency, which is the only one that depends on both $\rho$ and $w$. In fact, for a given energy, and since we have observed that the combinations of sines and cosines of $\phi_1$ in the probability vary slowly with $w$ (not shown here), each isocontour approximately corresponds to a constant value of
$w \sqrt{ \frac{c_0^2}{E_\nu^2} + \frac{2 c_0 c_1 \rho \cos 2\theta_{12}}{E_\nu} + c_1^2 \rho^2 }$, where $c_0 \equiv \Delta m_{21}^2/4$ and $c_1 \equiv -G_F y_e N_\text{Av} \cos^2\theta_{13} / \sqrt{2}$. Table \ref{TblChi0Conts} shows the minimum value of $\rho_\text{cavity}$ needed to reach $5\sigma$ and $10\sigma$ separations, for small ($w = 50$ km), medium ($100$ km), and large ($250$ km) cavities, after detector exposure times of $3$ months and $1.5$ years.

\begin{figure}[t!]
 \begin{center}
  \scalebox{0.17}{\includegraphics{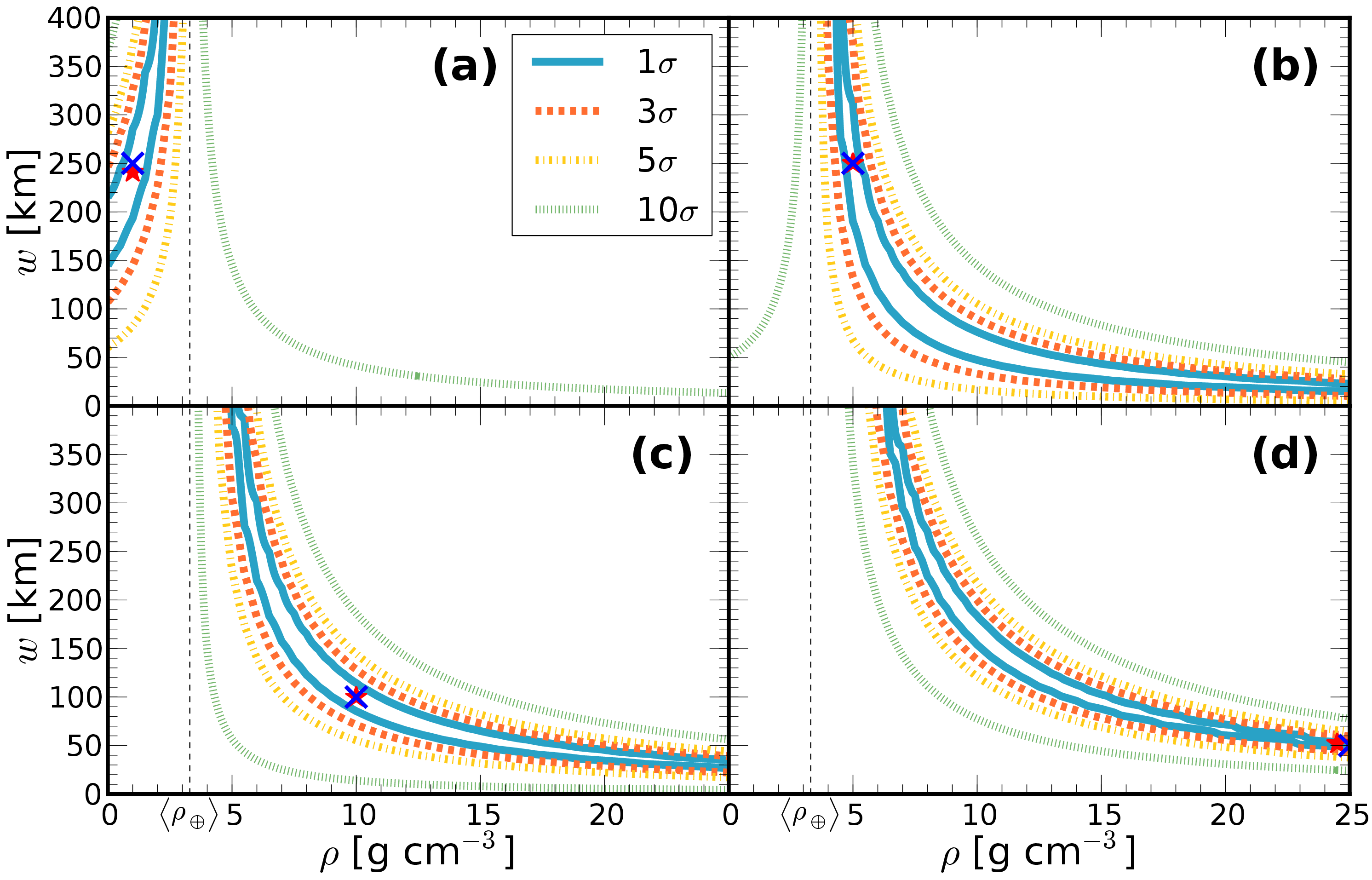}}
  \caption{(Color online) Confidence levels in the $w$ vs.~$\rho$ plane for an exposure time of 1.5 years. Plots (a)\---(d) correspond to baseline-centered cavities A\---D from Fig.~\ref{Fig-w.vs.rho}. The real values $\left(\rho_\text{0},w_0\right)$ are marked by crosses and the best-fit values by stars.}
  \label{Fig-w.vs.rho-rec}
 \end{center}
\end{figure}

\section{Determination of cavity parameters}

The next step in our analysis is to change the assumption of no-cavity to that of a real cavity, and to find the cavity's position, width, and density. This implies replacing $N_i^\text{no-cav} \to N_i^\text{cav}\left(w_0,d_0,\rho_\text{cavity}\right)$ in Eq.~(\ref{chi2.nocavity}). Heretofore, we set $1.5$ years as exposure time, unless otherwise specified, and keep the other details of the study equal to those presented in the sensitivity analysis of the previous section.

Firstly, we consider a cavity of known position, {\it i.e.}, centered on the baseline ($d = d_0 = (L_0-w)/2$) and find its width and density. We will study the four points marked in Fig.~\ref{Fig-w.vs.rho}: A ($\rho_\text{cavity} = 1$ g cm$^{-3}$, $y_e = 0.555$, $w_0 = 250$ km), corresponding to a cavity with an equivalent water-like density, motivated by Ref.~\cite{Pearson2014}; B ($\rho_\text{cavity} = 5$ g cm$^{-3}$, $y_e = 0.5$, $w_0 = 250$ km), to an iron-banded formation \cite{Morey:1983}; C ($\rho_\text{cavity} = 10$ g cm$^{-3}$, $y_e = 0.5$, $w_0 = 100$ km), to a heavier mineral deposit; and D ($\rho_\text{cavity} = 25$ g cm$^{-3}$, $y_e = 0.5$, $w_0 = 50$ km), representing a zone of seismic faults with the typical charge accumulation that supposedly exists prior to an earthquake of magnitude 7 in the Richter scale \cite{Pulinets:2004}. For cavity D, note that, in Ref.~\cite{Wang:2010cb}, a value of $\rho_\text{cavity} \approx \langle \rho_\oplus \rangle$ and a maximum value of $y_e \approx 4$ at the fault are used, while, to 
keep in line with the analysis of cavities A\---C, we have equivalently taken for cavity D $\rho_\text{cavity} = 25$ g cm$^{-3}$ and $y_e = 0.5$.

Notice that the closest distance from the cavities to the Earth surface, for A and B, is about $19$ km, and, for cavities C and D, $34$ km and $39$ km, respectively. Wider or uncentered cavities would lie closer to the surface; for instance, a baseline-centered cavity with $w_0 = 323$ km would lie only $12$ km deep, which is roughly the current maximum drilling depth \cite{Fuchs:1990}.

In Fig.~\ref{Fig-w.vs.rho-rec} we observe that values of $\rho$ close to $\langle\rho_\oplus\rangle$ (equivalent to the no-cavity case), regardless of the value of $w$, are at the same significance level as shown in Fig.~\ref{Fig-w.vs.rho}. The shape of the isocontours is explained by the same argument as in said figure. Notice that the uncertainty in $w$, for a fixed $\rho$ hypothesis, decreases as the real cavity density gets farther away from the Earth's mean density. Furthermore, the allowed ranges of values of $\rho$ and $w$ are large in all cases, which indicates that more information is needed to determine the cavity parameters. In this sense, it is interesting to point out that in a real\--case scenario, where either there could be some prior knowledge about the density or a particular cavity density is being searched for, it would be possible to constrain $w$ significantly. For instance, for $\rho = 1$ g cm$^{-3}$ (case A), $w = 240_{-50}^{+30}$ km at $1\sigma$ uncertainties, while for $\rho = 25$ g cm$^{-3}$ (case D), $w = 50\pm10$ km. 

\begin{figure}[t!]
 \begin{center}
  \scalebox{0.17}{\includegraphics{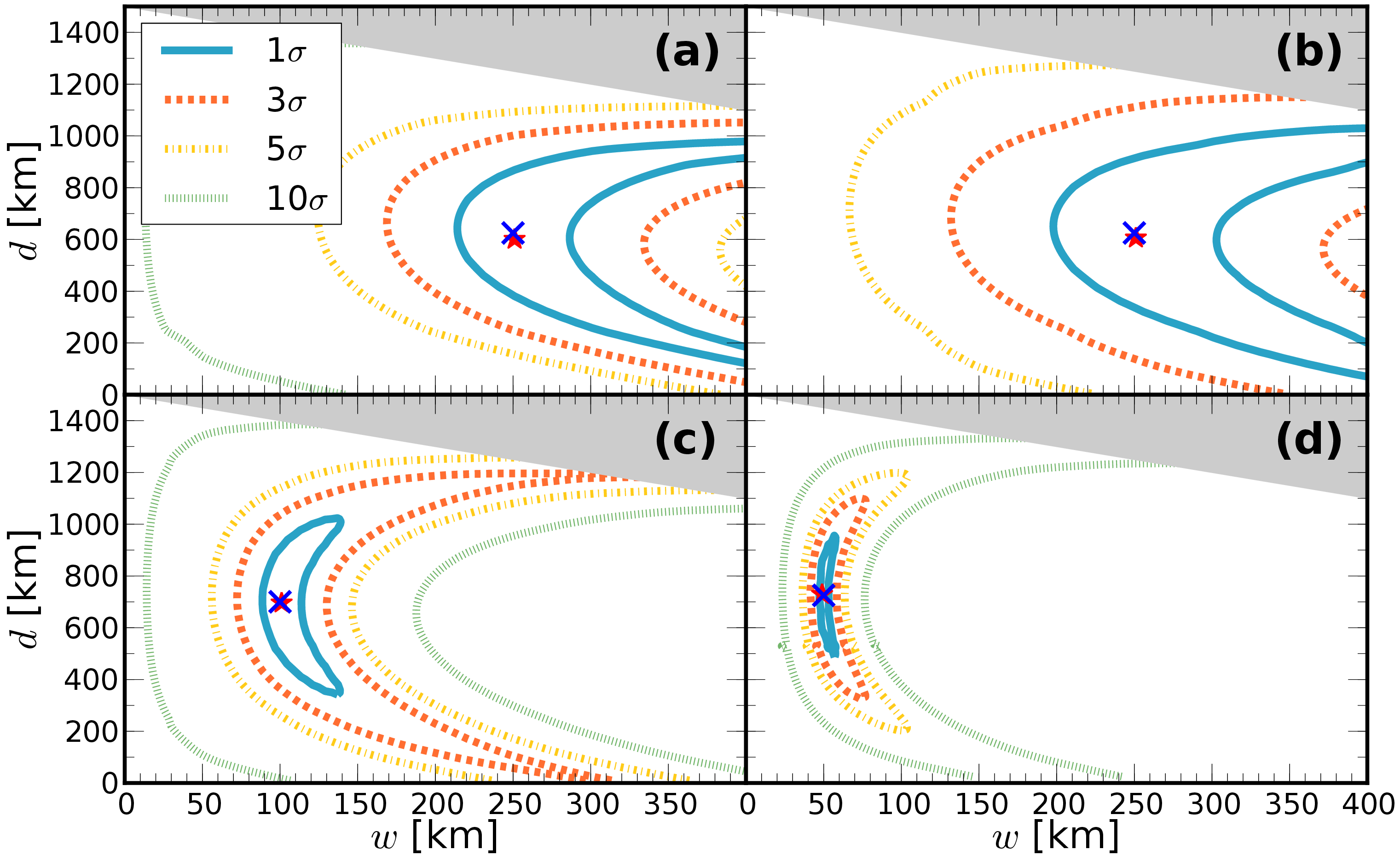}}
  \caption{(Color online) Confidence levels in the $d$ vs.~$w$ plane for an exposure time of $1.5$ years. Plots (a)\---(d) correspond to baseline-centered cavities A\---D from Fig.~\ref{Fig-w.vs.rho}. The real values $\left(w_0,d_0\right)$ are marked by crosses and the best-fit values by stars.}
  \label{Fig-d.vs.w}
 \end{center}
\end{figure}

Secondly, motivated by the discussion at the end of the preceding paragraph, we consider the case where the density of the cavity is known and find its position and width. In Fig.~\ref{Fig-d.vs.w}, we show the contour regions of the four cavity cases A\---D in the $w$ and $d$ plane (with $w+d \leq L_0$). In this figure, the sizes of the $w$ allowed regions are much smaller than in the previous analysis in the $w$ vs.~$\rho$ plane. Therefore, we are capable of determining $w$ with reasonable precision in cases C and D. On the other hand, the determination of $d$ improves mildly as the density increases. For example, in case A, $d = 600_{-475}^{+375}$ km and, in D, $d = 725_{-275}^{+225}$ km at $1\sigma$ uncertainties. This is because the assumption of a known cavity density has a greater impact in the oscillation probability than knowing the cavity position. In fact, using the slab approximation, the weak dependence of the probability on $\phi_1$ and $\phi_3$, and, therefore, on $d$, for most of the relevant energy range, is clear.

\begin{figure}[t!]
 \begin{center}
  \scalebox{0.48}{\includegraphics{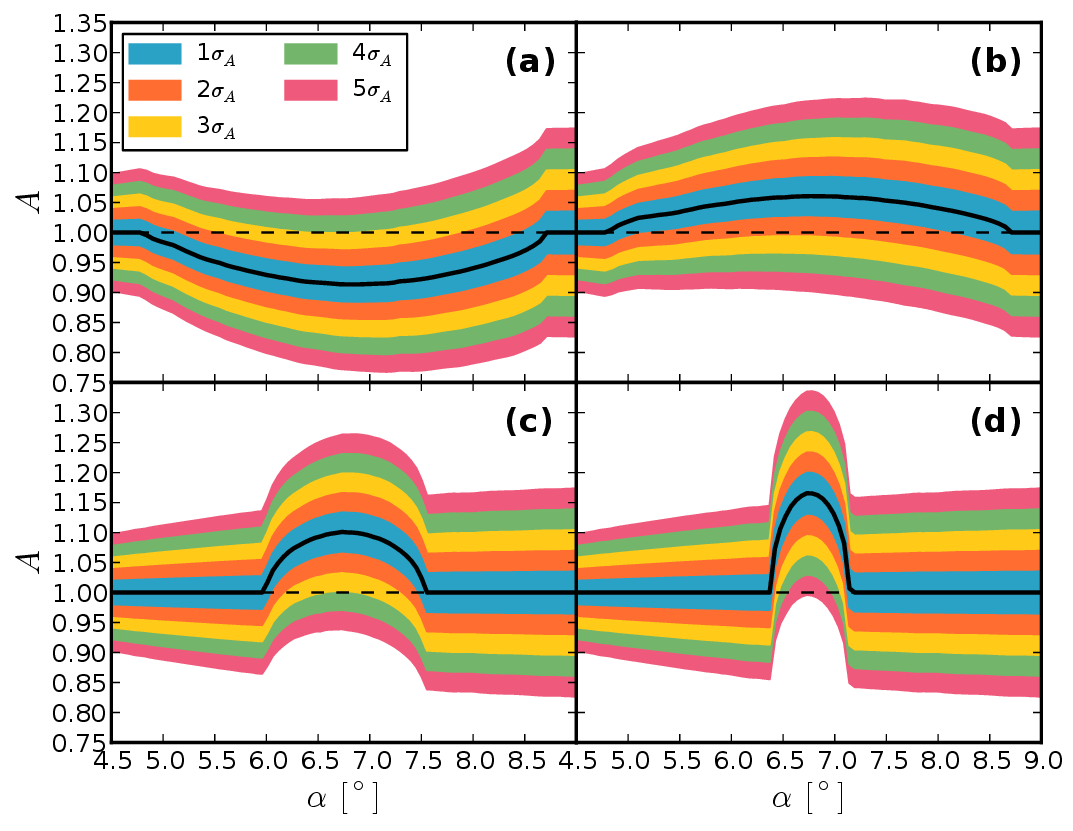}}
  \caption{(Color online) In solid black lines, $A$ vs.~$\alpha$, for the four cavities A\---D, together with selected uncertainty regions. The exposure time is $3$ months.}
  \label{Fig-ACCC}
 \end{center}
\end{figure}

\section{Searching for a cavity}

We have explored the possibility of varying the beam orientation, defined by the angle $\alpha$ measured with respect to the tangent to Earth at S, as a way of finding a cavity. This is shown in Fig.~\ref{FigCavityConfiguration}b, and could be implemented by using a mobile neutrino detector, either on land \cite{DeRujula:1983ya} or at sea \cite{Huber:2009kx}. At an angle $\alpha$, the beam travels a baseline $L_\alpha$; for some values of $\alpha$, it will cross a portion of the cavity with position $d_\alpha$ and width $w_\alpha$. To quantify the size of the traversed portion, we have defined
\begin{equation}\label{EqDefA}
 A\left(L_0,w_0,d_0,\alpha,\rho_\text{cavity}\right) 
 = \frac{N^\text{cav}\left(L_\alpha,w_\alpha,d_\alpha,\rho_\text{cavity}\right)}{N^\text{no-cav}\left(L_\alpha\right)} ~,
\end{equation}
with $N^\text{cav}$ and $N^\text{no-cav}$ the total numbers of $e^+$ between $5$ and $150$ MeV, and $L_\alpha$, $w_\alpha$, $d_\alpha$ functions of $L_0$, $w_0$, $d_0$, and $\alpha$. Deviations from $A=1$ indicate the presence of a cavity. For this analysis, we have set the exposure time at $t = 3$ months.

Fig.~\ref{Fig-ACCC} shows the $A$ vs.~$\alpha$ curves for the four cavities A\---D. The curve for cavity A lies below $A=1$ because its density is lower than $\langle\rho_\oplus\rangle$, while the densities of cavities B\---D are higher. The $1\sigma_A$ to $5\sigma_A$ uncertainty regions around the curves are included, with $\sigma_A \equiv A \sqrt{\left(1/N^\text{cav}\right)\left(1+A\right)}$ the standard deviation of $A$. Since $N^\text{cav} \sim 1/L_\alpha^2$ (see the definition of $\Phi_\text{tot}$ following Eq.~(\ref{equ:DefNumberPositrons})), then $\sigma_A \sim L_\alpha$ and, given that, by geometrical construction, $L_\alpha$ grows with $\alpha$ (as long as $\alpha < 90^\circ$), this means that $\sigma_A$ also grows with $\alpha$, a feature that is observed in Fig.~\ref{Fig-ACCC}. Similarly to previous figures, the separation from $A = 1$ grows as $\rho_\text{cavity}$ moves away from $\langle\rho_\oplus\rangle$. The C.L. achieved at the maximum deviation point is not as high as those shown 
in Fig.~\ref{Fig-w.vs.rho}, due to the fact the analysis of $A$ does not take into account the spectral shape, but only the total 
$e^+$ count at the detector.

Separation from $A=1$ at the $2\sigma_A$ C.L. is achieved for all cavities, with A and C reaching $3\sigma_A$, and D reaching $5\sigma_A$. Thus, the parameter $A$ could be used as a quick estimator of the presence of a cavity, while a more detailed analysis, similar to the one performed for Fig.~\ref{Fig-d.vs.w}, could be used to estimate the cavity shape, {\it i.e.}, its position and width, and, from the latter, its volume. 

\section{Conclusions}

We have studied the use of a low-energy ($5$\---$150$ MeV) $\beta$-beam of $\bar{\nu}_e$, with a baseline of $1500$ km and a large luminosity enhancement, to find the presence of deep underground cavities in the Earth's crust. In the context of a more detailed analysis than prior work, we have determined for the first time the sensitivity of the experimental setup as a function of the cavity density $\rho$ and size $w$ (dimension of the cavity aligned with the neutrino beamline), which reaches significances in the order of $5 \sigma$ ($3 \sigma$) for baseline-centered cavities with densities lower than $1$ g cm$^{-3}$ or greater than $5$ g cm$^{-3}$, exposure time of $1.5$ years ($3$ months), and $w$ greater than $200$ km. As a result, we have elaborated a roadmap that can be used to assess the possibility of discovering a cavity with an arbitrary density, at a high confidence level, which increases as the density of the cavity differs more from that of the surrounding Earth's crust. We analyzed the C.L. regions of the reconstructed parameters of four different cavities in the $\rho$ vs.~$w$ plane, for a known cavity position $d$, and also in the $w$ vs.~$d$ plane, when $\rho$ is known. In general, the allowed regions are large, but when there is knowledge of $\rho$ the uncertainty on the cavity width is dramatically reduced, {\it e.g.}, for $\rho =1$ g cm$^{-3}$, the water equivalent case, and for $\rho =25$ g cm$^{-3}$, the seismic fault scenario, the cavity width can be determined to within 20\% error. Unfortunately, the uncertainty in the cavity position is always large, {\it e.g.}, for $\rho =1$ g cm$^{-3}$ ($25$ g cm$^{-3}$) it has an error of 80\% (40\%). 

It should be noted that there is an intrinsic triple degeneracy in the proposed method, between the position, dimensions, and density of the cavity. This can be evidenced in Fig.~\ref{Fig-w.vs.rho-rec}, which shows that many different combinations of $w$ and $\rho$ can fit a given detector signal comparably well. Simply put, it is equally valid to interpret the signal at the detector as having been generated by a small cavity with high density, or by a large cavity with lower density; ignorance of the position of the cavity adds an extra degree of freedom. Independent knowledge of one or two of these parameters could be used to break the degeneracy, either partially or totally.

Finally, we have considered sweeping the Earth in search of a cavity using an orientable neutrino beam. In order to do this, we have implemented a rate ratio analysis, which proves to be a useful tool to detect the presence of a cavity, with at least $2\sigma$ statistical significance, and reaching up to $5\sigma$ for high densities ($25$ g cm$^{-3}$) or, equivalently, lower densities and higher electron fractions.

\begin{acknowledgments}
The authors would like to thank the Direcci\'on de Inform\'atica Acad\'emica at the Pontificia Universidad Cat\'olica del Per\'u (PUCP) for providing distributed computing support in the form of the LEGION system, and Teppei Katori, Arturo Samana, Federico Pardo-Casas, and Walter Winter for useful information and feedback. This work was funded by the Direcci\'on de Gesti\'on de la Investigaci\'on at PUCP through grant DGI-2011-0180.
\end{acknowledgments}

\end{document}